\documentclass[aps,prb,twocolumn,groupedaddress]{revtex4}

\usepackage{graphicx}
\usepackage{subfigure}

\def\be{\begin{equation}}
\def\ee{\end{equation}}
\def\bea{\begin{eqnarray}}
\def\eea{\end{eqnarray}}

\begin{document}


\title{ Renormalization group method for weakly coupled quantum chains:
comparison with exact diagonalization }


\author{J.V. Alvarez$^1$ and S. Moukouri$^{1,2}$}
\affiliation{$^1$Department of Physics, University of Michigan,
             $^2$Michigan Center for Theoretical Physics
         2477 Randall Laboratory, Ann Arbor  MI 48109}


\date{\today}

\begin{abstract}
We show that numerical quasi-one-dimensional renormalization group  
allows accurate study of weakly coupled chains 
with modest computational effort.
We perform a systematic comparison with exact diagonalization 
results in two and three-leg spin ladders with a transverse Hamiltonian
that can involve frustration. 
Due to the variational nature of the algorithm,
the accuracy can be arbitrarily improved enlarging the basis 
of eigenstates of the density matrix defined in the transverse direction.
We observe that the precision of the algorithm 
is directly correlated to the binding of the chains.    
We also show that the method performs especially well in frustrated 
systems.    
\end{abstract}

\maketitle


\section{Introduction}
It was recently shown \cite{MAT-moukouri} that a general Kato-Bloch
matrix expansion can be applied to weakly-coupled quantum chains.
This algorithm was used to study weakly-coupled Heisenberg chains
\cite{TS1-moukouri,TS2-moukouri}. The DMRG was used as the method of
solution for an isolated chain and then again for the solution of an
effective 1D model which is obtained by projecting the problem
to the basis of the tensor product of independent chain states.
A good agreement with the stochastic series expansion (SSE) quantum
Monte Carlo (QMC) was found for transverse couplings $J_{ \perp}$ not
too large. Then interchain diagonal exchange $J_d$ which frustrates 
the system was introduced.  It is found, by analyzing ground state
energies and spin-spin correlation functions, that there is a transition
between two ordered magnetic states. When $J_{d}/J_{\perp} \alt 0.5$,
the ground state displays a N\'eel order. When
$J_{d}/J_{ \perp} \agt 0.5$, a collinear magnetic ground state
in which interchain spin correlations are ferromagnetic becomes stable.
In the vicinity of the transition point, $J_{d}/J_{ \perp} \approx 0.5$,
the ground state is disordered. The prediction of a disordered ground
state is of central importance for two reasons.  First, because a recent
neutron scattering experiment \cite{Coldea} on the frustrated AFM
Cs$_2$CuCl$_4$ has predicted a spin liquid ground state in this material.
 Second, a disordered doped spin liquid has been conjectured to be relevant for
the physics of high temperature cuprate superconductors. The search of
this disordered two-dimensional state by numerical methods has been 
challenging.
Cluster QMC methods \cite{EVERTZ,SANDVIK},
that have been extremely useful in nonfrustrated spin systems,     
are hampered by sign problems for
Hamiltonians with finite $J_{d}$, making very difficult their 
study by this technique.  
New algorithms have been specifically designed to deal with frustration 
and intense numerical
research has been devoted to these systems \cite{DAGOTTO,JONGH,CAPRIOTTI}. It
is thus of central importance to show that the DMRG prediction is
correct. 

We address in this paper several questions that 
increase the understanding and show new  potential of the method,
giving additional support for the physical findings 
of Ref. \onlinecite{TS2-moukouri}.
Some of these questions are technical in nature and they demand 
exhaustive comparison with exact results.

\begin{figure}
\includegraphics[width=6.0cm]{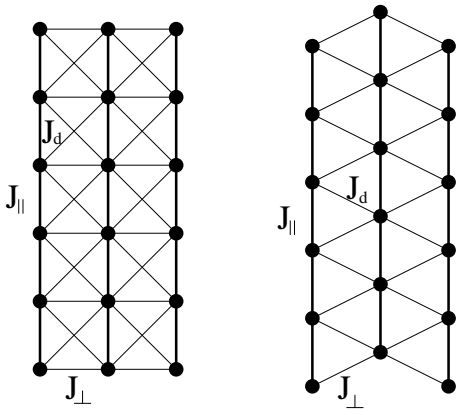}
\caption{The two lattice geometries studied in the text.
On the left, chains coupled with perpendicular $J_{\perp}$
and diagonal coupling $J_{d}$. On the left the chains 
are arranged on a triangular lattice. 
  \label{ED-latt}}
\end{figure}

The class of models that we study here have transverse terms 
involving competing interactions. These transverse terms are 
projected in a optimal reduced basis of eigenstates  of the
independent chain. If, for instance, the chains are coupled 
with perpendicular and diagonal exchange constants 
(see the left ladder in Fig. \ref{ED-latt}),   
the projection of the perpendicular ($J_{\perp}$)
and the diagonal ($J_{d}$) parts of the interchain coupling
is qualitatively different one to the other.
More precisely, to represent coupling terms along the diagonal
requires  matrix elements of operators defined in different sites
(those usually associated with the computation
of short range correlation functions in 1D DMRG calculations). 
It is this competing behavior what generates negative local 
Boltzmann weights (sign problems) that 
can not be eliminated by canonical transformations 
when QMC is used. Therefore, it is important to
check if all competing terms are represented  and treated
with similar accuracy by studying models 
that mix these terms in different ways.

We will also show that the accuracy is directly correlated to the 
binding energy of the chains and not to the nominal 
values of the transverse couplings, concluding that the method is specially 
good for the study of frustrated systems.  
In addition, based on the results of this study we have designed 
internal tests that signal good performance of the method
when dealing with larger lattices where comparison 
with exact data is not possible.

Finally, we are also interested in showing the   
controlability of the approximation involved in the method, 
specifically for frustrated
systems. We emphasize that its variational nature implies   
that the accuracy can be systematically improved by enlarging 
the number of states kept in the density matrix defined perpendicular 
to the chains($m_{s2}$). This point is not a trivial one; it demands an 
accurate projection of the transverse Hamiltonian in an accurate 
representation of the Hilbert space of the chains.  
The  systematic comparison with exact diagonalization (ED) results
in two- and three-legged ladders presented here shows that in Q1D 
systems excellent results can be obtained 
with modest values for $m_{s2}$. 

The rest of the paper is organized as follows. 
The different models studied in 
this paper are presented in section \ref{Models}. 
Then we summarize the main steps of the algorithm implementation
in section \ref{Method}.  In section \ref{ms1ms2}
we show how to improve systematically the numerical 
results for several magnitudes  
exploiting the variational property of the method.    
In section \ref{Frust} we study the accuracy  as we 
increase different exchange couplings in the transverse 
Hamiltonians. In section \ref{conclusions} 
we present our conclusions.
    
\section{Models}
\label{Models}
To discuss the issues that we presented in the introductory
section  we need to compare the results obtained with 
two-step DMRG with exact data,
that can be achieved by ED in small lattices 
\cite{BARNES}. In addition, 
we want to evaluate the performance of the method 
for a very general class of transverse Hamiltonians 
including those that involve  frustrating couplings between the chains
( i.e. , an  exchange constant $J_{d}$ along the diagonal 
of the square lattice). It is also important to study systems with different numbers of legs
because they behave in a very different way as they approach the thermodynamic
limit \cite{SCIENCE-LAD}. 

The natural choice as test models are then
the Heisenberg two- or three-legged ladders in the strong coupling limit: 
\begin{eqnarray*}
&&H=H_{\parallel}+H_{\perp}+H_{d} \\
&&H_{\parallel} =J_{\parallel}\sum_{j,l}{\bf S}_{j,l}{\bf S}_{j+1,l} \\
&&H_{\perp}=J_{\perp} \sum_{j,l}{\bf S}_{j,l}{\bf S}_{j,l+1} \\
&&H_{d}=J_{d} \sum_{j,l}({\bf S}_{j,l}{\bf S}_{j+1,l+1}+
{\bf S}_{j+1,l}{\bf S}_{j,l+1})\\
\label{square}
\end{eqnarray*}
\noindent 
where $J_{\parallel}>J_{\perp},J_{d}$.
${\bf S}_{j,l}$ represents the spin operator in the 
site $j$ of chain $l$ with the indices running  
$i=j \ldots L_{x}$ and $l=2,3$. We impose open
boundary conditions in the two directions.  
The coupling $J_{d}$ connects spin operators along the diagonal 
and introduces frustration in the model.

Another frustrated model of great interest is the system of weakly coupled 
chains arranged in a anisotropic triangular lattice which can be written 
in the form
\begin{equation}
H=H_{\parallel}+H_{\perp}+H'_{\perp}
\label{HMATL}
\end{equation}
where the new term $H'_{\perp}$ is:
\begin{equation} 
H'_{\perp}= J_{d}\sum_{j,l}{\bf S}_{j,l}{\bf S}_{j+1,l+1}
\end{equation}
\noindent 

Note that the model in the square lattice with diagonal couplings 
(\ref{square}) reduces to the model defined in the triangular
lattice \ref{HMATL} when one of the two diagonal couplings
is set to zero.
 
 \section{Method and data analysis}
\label{Method}
 A detailed description of the two-step DMRG method, which we will refer 
for short 2S-DMRG, has been given in Refs.
\onlinecite{TS1-moukouri} and \onlinecite{TS2-moukouri}. 

The exact spectrum of a single AF chain is known from the Bethe
ansatz, but eigenfunctions are not easily accessible. Thus,
the density-matrix renormalization group (DMRG) method \cite{WHITE,DMRG-book}
will be used to compute an approximate spectrum $\epsilon_n, |\phi_n \rangle$
of a single chain.  A preliminary account of this approach
\cite{TS1-moukouri} as well as an extensive comparison with the
Quantum Monte Carlo method was presented elsewhere \cite{TS2-moukouri}.
By expressing the Hamiltonian on the basis generated
by the tensor product of the states of different chains one obtains,
up to the second order, the effective one-dimensional Hamiltonian,

\begin{eqnarray}
\nonumber \tilde{H} \approx \sum_{[n]} E_{\parallel [n]} 
|\Phi_{\parallel[n]}\rangle \langle \Phi_{\parallel [n]}| +
 J_{\perp} \sum_{l} {\bf \tilde{S}}_{l} {\bf \tilde{S}}_{l+1}+\\
J_{d} \sum_{l} {\bf \tilde{S}}_{l}{\bf \tilde{S}}_{l+1}
+...
\label{efhamil}
\end{eqnarray}

\noindent where the composite chain-spin operators on the chain $l$ are
${\bf \tilde{S}}_{l}=({\bf \tilde{S}}_{1l},
 {\bf \tilde{S}}_{2l}, ...{\bf \tilde{S}}_{Ll})$,  $L_x$ is the
chain length. The matrix elements of the first  order local
spin operators are respectively

\begin{equation}
\nonumber {\bf \tilde{S}}_{i,l}^{n_l,m_l}=\langle \phi_{n_l}|{\bf S}_{i,l}|\phi_{m_l}\rangle
\end{equation}


We will study the following magnitudes. 
The ground state energy per site $E_G=E_0/L$, where $L=L_{x}L_{y}$, 
the gap to the first excited state 
in the sector of spin $S_z=0$, $\Delta_0$ and  $S_z=1$, $\Delta_1$. 
To monitor both gaps, that have to be equal in a spin rotational invariant
Hamiltonian, is a very astringent probe of the reliability of the method.
The interchain spin flip coupling terms 
(e.g. $J_\perp S_{i,l}^{+}S_{i,l+1}^{-}$+h.c.) connects different sectors 
with different total chain magnetization $S^{z}_{T}=\sum_{i}S^{z}_{i}$
that have to be accurately and consistently described by the 
$m_{s2}$ states outcoming from the block renormalization step that 
culminates the one-dimensional part of the algorithm.
This is achieved by a careful targeting of states in  sectors 
with increasing chain magnetization $S^{z}_{T}$. 
We also computed the binding energy of the chains 
$E_B=(E_0(J_{\perp},J_{d})-E_0(J_{\perp}=0,J_{d}=0))/L$ 
which is primarily a 'transverse' magnitude.  The ED data 
have been obtained using the Davidson algorithm \cite{DAVIDSON} 
and imposing open boundary conditions.    
\begin{table}
\caption{\label{tab:table1} Ground state energy per site ($E_{G}$), 
 gaps to the $S_z=0$ ($\Delta_0$) and $S_z=1$ ($\Delta_{1}$) 
 excitations and binding energy for a 2x10 leg-ladder with 
 $J_{\perp}=0.1$, $\rho=0.0$ as a function of the number 
of states kept in the parallel and perpendicular direction.}  
\begin{ruledtabular}
\begin{tabular}{|c|c|c|c|c|c|}
\multicolumn{2}{|c|}{2x10} &$E_G$&$\Delta_{0}$&$\Delta_{1}$&$E_B$ \\  \hline
\multicolumn{2}{|c|}{ED}&-0.4273296  & 0.283182   & 0.283182  & -.0015261  \\ \hline
$m_{s1}$&$m_{s2}$&$E_0$&$\Delta_{0}$&$\Delta_{1}$&$E_B$ \\ \hline
8&8 & -0.42433209  &0.48038304&0.28121407 & -.00083699      \\\hline
8&16& -0.42443195  &0.47999406&0.27884533 & -.00093685      \\ \hline
8&32& -0.42446616  &0.47825339&0.27731861 & -.00097105      \\\hline
16&8& -0.42676461  &0.29092770&0.28842390 & -.00098189      \\\hline
16&16&-0.42698277  &0.29235105&0.28683188 & -.00120005      \\\hline
16&32&-0.42707158  &0.28790788&0.28491220 & -.00128885      \\\hline
16&48&-0.42712096  &0.28737740&0.28416394 & -.00133824      \\\hline
16&64&-0.42712439  &0.28674689&0.28360519 & -.00134167      \\\hline
32&8 &-0.42679647  &0.28862051&0.28861954 & -.00099295      \\\hline
32&16&-0.42703063  &0.29022064&0.29021987 & -.00122710      \\\hline
32&32&-0.42716088  &0.28579872&0.28610427 & -.00135735      \\\hline
32&48&-0.42724324  &0.28531525&0.28552398 & -.00143972      \\\hline
32&64&-0.42725807  &0.28391776&0.28397001 & -.00145454      \\\hline 
32&96&-0.42729144  &0.28342612&0.28342613 & -.00150668               
\end{tabular}
\end{ruledtabular}
\end{table}

\begin{table}
\caption{\label{tab:table2} Same magnitudes that in 
table \ref{tab:table1} in the 2x10 frustrated ladder with 
$J_{\perp}=0.1$ $\rho=0.5$ } 
\begin{ruledtabular}
\begin{tabular}{|c|c|c|c|c|c|}
\multicolumn{2}{|c|}{2x10} &$E_G$&$\Delta_{0}$&$\Delta_{1}$&$E_B$ \\  \hline
\multicolumn{2}{|c|}{ED}&-.42593175  &0.319339   &0.319339   &-0.00012825   \\ \hline
$m_{s1}$&$m_{s2}$&$E_0$&$\Delta_{0}$&$\Delta_{1}$&$E_B$ \\ \hline
8&8 &-.42352572 & 0.49158956&  0.31988540 &  -.00003061       \\\hline
8&16&-.42353999 & 0.49158613&  0.31989184 &  -.00004488      \\ \hline
8&32& -.42354330&  0.49138272&  0.31979321&  -.00004819       \\\hline
16&8&  -.42581768&  0.32153617&  0.31961081&   -.00003496      \\\hline
16&16& -.42584056&  0.32175918&  0.31979066&   -.00005784     \\\hline
16&32& -.42587346&  0.32188676&  0.32002579&   -.00009073     \\\hline
16&48& -.42587597&  0.32147706&  0.31966604&   -.00009325     \\\hline
16&64& -.42587630&  0.32141332&  0.31957621&   -.00009357     \\\hline
32&8 & -.42583981&  0.31970010&  0.31969943&   -.00003628     \\\hline
32&16&  -.42586676&  0.31986620&  0.31986620&  -.00006323    \\\hline
32&32&  -.42591964&  0.32024093&  0.32027350&  -.00011612   \\\hline
32&48&  -.42592399&  0.31979512&  0.31995252&  -.00012046   \\\hline
32&64&  -.42592794&  0.31952783&  0.31952783&  -.00012441    \\\hline
32&96&  -.42593076&  0.31936894&  0.31933910&  -.00012724                      \end{tabular}
\end{ruledtabular}
\end{table}

\section{2S-DMRG: a controlled approximation}
\label{ms1ms2}
The main purpose of this section is 
to compare the efficiency of the algorithm in frustrated and unfrustrated 
systems. Here we keep all the exchange couplings in the model  constant 
and we show how the precision of the method improves 
systematically as we  increase the number of states 
of the density matrix in the direction
parallel ($m_{s1}$) and perpendicular to the chains ($m_{s2}$).
The parallel coupling $J_{\parallel}=1.0$ as the reference scale 
and we will keep it fixed as the reference energy scale in the 
problem.  
We present an exhaustive comparison of the 2x10 and the 
3x6 ladders taking a common value of the perpendicular exchange coupling
$J_{\perp}=0.1$.  To study the role of frustration we take two values  
of $J_{d}=\rho J_{\perp}$ corresponding with $\rho=0$, 
tables \ref{tab:table1} (2x10) and \ref{tab:table3} (3x6) and 
$\rho=0.5$, tables \ref{tab:table2} (2x10) and \ref{tab:table4}(3x6),      
for different number of states kept 
in both directions.    
Since the Hamiltonians are anisotropic most of the energy 
is stored along the chains 
in the $J_{\parallel}$ bonds 
and therefore we observe larger increments in accuracy  
when we increase $m_{s1}$ at fixed $m_{s2}$ than when we increase  $m_{s2}$ 
at fixed $m_{s1}$. 
It is also clear from this study that in all cases the singlet-triplet 
gap $\Delta_{1}$
is closer to the exact value obtained with ED than the gap in the 
$S^{T}_{z}=0$ sector $\Delta_0$. 
This is a consequence of the variational
nature of the method and it is also observed 
in the conventional 1D DMRG: $\Delta_{1}=E_0(S=1)-E_0(S=0)$ involves two 
minimal energies in different sectors and it is computed 
with more precision than  $\Delta_{0}=E_1(S=0)-E_0(S=0)$.
However, the difference between the two gaps vanishes as 
$m_{s1}$ and $m_{s2}$ increase. Similar dependence in 
$m_{s1}$ and $m_{s2}$ is found in the anisotropic triangular lattice.
 
We also observe that, both in the two- and the three-leg ladder, 
the accuracy is always higher in the frustrated case. Actually, the 
improvement is one order of magnitude in all 
the quantities computed, irrespective of the values of  
$m_{s1}$ and $m_{s2}$.


\begin{table}
\caption{\label{tab:table3} 
Energies and gap dependence on the number of states kept in the 3x6 
ladder with $J_{\perp}=0.1$, $\rho=0.0$.}
\begin{ruledtabular}
\begin{tabular}{|c|c|c|c|c|c|}
\multicolumn{2}{|c|}{3x6} &$E_G$&$\Delta_{0}$&$\Delta_{1}$&$E_B$\\  \hline
\multicolumn{2}{|c|}{ED}  &-0.41733967 &0.423312 &0.423312 &-0.001743481\\\hline
$m_{s1}$&$m_{s2}$&$E_G$&$\Delta_{0}$&$\Delta_{1}$&$E_B$ \\ \hline
4&4 & -.36497451 &  0.32915698 & 0.16977127 & -.00256288    \\  \hline
4&8 & -.36518753 & 0.32932855 &  0.17029788 & -.00277591    \\   \hline
4&16& -.36521796 & 0.32976460 & 0.16657322  & -.00280634    \\   \hline 
8&4 & -.41690780 & 0.44817021 & 0.44817019  & -.00111016    \\   \hline
8&8 & -.41692515 & 0.42847412 & 0.42847414  & -.00132896    \\   \hline
8&16& -.41723072& 0.42941491 & 0.42941433   & -.00163453    \\   \hline
8&32& -.41733685 & 0.42386884 & 0.42386885  & -.00174066  \\   
\end{tabular}
\end{ruledtabular}
\end{table}

\begin{table}
\caption{\label{tab:table4} 
Energies and gaps as a function of the  number of states kept in the 3x6 
ladder with $J_{\perp}=0.1$ and frustration $\rho=0.5$.}
\begin{ruledtabular}
\begin{tabular}{|c|c|c|c|c|c|}
\multicolumn{2}{|c|}{3x6} &$E_G$&$\Delta_{0}$&$\Delta_{1}$&$E_B$\\  \hline
\multicolumn{2}{|c|}{ED}  &-0.41580272 &0.47413 &0.47413&-0.00020653 \\  \hline
$m_{s1}$&$m_{s2}$&$E_G$&$\Delta_{0}$&$\Delta_{1}$&$E_B$  \\ \hline
4&4 &   -.36273211 &  0.31690894  & 0.21046881 & -.00032049  \\  \hline
4&8 &   -.36274150 &  0.31664021  & 0.21032338 & -.00032987  \\ \hline
4&16&   -.36521796 &  0.32976460  & 0.16657322 & -.00280634  \\ \hline 
8&4 &   -.41565327 &  0.47451007  & 0.47451008 & -.00005708   \\ \hline
8&8 &   -.41572257 &  0.47473022  & 0.47473022 & -.00012638  \\ \hline
8&16&   -.41579631 &  0.47495257  & 0.47495259 & -.00020012  \\ \hline
8&32&   -.41580249 &  0.47415896  & 0.47415897 & -.00020630  \\ 
\end{tabular}
\end{ruledtabular}
\end{table}


\section{Performance in frustrated lattices.} 
\label{Frust}  
\subsection{Chains coupled with perpendicular and diagonal couplings}

To study this point in more detail we have computed all the magnitudes 
listed above  as a function of $\rho=J_{d}/J_{\perp}$. 
the results are presented in table \ref{tab:table5}  for the 2x10 system 
and in table \ref{tab:table6}  for the 3x6 system. 
Both $E_G$ and $E_B$ are non-monotonic functions of $\rho$
with extrema in the vicinity of $\rho=0.6$.  Actually, $E_G(\rho)$ 
is a function concave up and $E_B(\rho)$ is concave down. 
The maximum in the $E_{B}$ is the remnant 
of the decoupling between staggered parts of the 
coarse-grained spin operator predicted 
by field theory \cite{ALLEN} \cite{TSVELIK}
at $\rho=0.5$.  
In finite open-boundary systems the 'decoupling point' takes place at higher
values of  $\rho$  but still can be observed 
through the decrease of the binding energy of the chains\cite{TS2-moukouri}.
The error in the ground state energy 
is reduced when $\rho$ increases, making clear that 2S-DMRG 
can deal with coupling Hamiltonians running along the diagonal 
and even more, it can perform better with frustrated than with 
nonfrustrated systems.  A crucial observation here 
is that the ground state error $\delta E_0=(E_0-E_{0}^{\text{ED}})/L$ 
does not increase with $J_{d}$ but   
exhibits nearly linear correlation with the value of $E_B$ i.e 
it also has a minimum at $\rho \sim 0.6$.  
As we are going to see, this  widens remarkably the range of  
applicability of the method.  

\begin{table}
\caption{\label{tab:table5} 
Energy and gap $\Delta_1$ as a function of the ratio 
$\rho=J_{\perp}/J_{\parallel}$ for the 2x10 ladder with 
$J_\perp=0.1$.}
\begin{ruledtabular}
\begin{tabular}{|c|c|l|l|l|}
$2\times10$&$\rho$ &$E_G$&$\Delta_{1}$&$E_B$         \\ \hline
ED    &0.1&-0.4268910 & 0.288913   &-.0010875   \\ \hline
2S-DMRG&0.1&-0.4268418 & 0.28952981 &-.00103835  \\ \hline
ED    &0.2&-0.4265297 & 0.295369   &-.0007262   \\ \hline
2S-DMRG&0.2&-0.4264985 & 0.29583873 &-.00069499  \\ \hline
ED    &0.3&-0.4262480 & 0.302577   &-.0004445   \\ \hline
2S-DMRG&0.3&-0.4262304 & 0.30292659 &-.00042692  \\ \hline
ED    &0.4&-0.4260482 & 0.310563   &-.0002447   \\ \hline
2S-DMRG&0.4&-0.4260397 & 0.31081613 &-.00023620  \\ \hline 
ED    &0.5&-0.4259317 & 0.319339   &-.0001282   \\ \hline 
2S-DMRG&0.5&-0.4259279 & 0.31952206 &-.00012441  \\ \hline
ED    &0.6&-0.4258999 & 0.325816   &-.0000964   \\ \hline
2S-DMRG&0.6&-0.4258961 & 0.32594461 &-.00009263  \\ \hline
ED    &0.7&-0.4259532 & 0.315247   &-.0001497   \\ \hline
2S-DMRG&0.7&-0.4259449 & 0.3153660  &-.00014140  \\ \hline
ED    &0.8&-0.4260921 & 0.305058   &-.0002886   \\ \hline
2S-DMRG&0.8&-0.4260743 & 0.30519779 &-.00027075  \\ \hline
ED    &0.9&-0.4263161 & 0.295269   &-.0005126   \\ \hline
2S-DMRG&0.9&-0.4262837 & 0.29546053 &-.00048020  \\ \hline
ED    &1.0&-0.4266246 & 0.285893   &-.0008211   \\ \hline
2S-DMRG&1.0&-0.4265723 & 0.286168   &-.00076881  \\ 
\end{tabular}
\end{ruledtabular}
\end{table}
The error in the gap $\delta \Delta_1=|\Delta_1-\Delta_{1}^{\text{ED}}|$ 
follows a similar trend. However, the value of $\delta \Delta_1(\rho)$ 
has a minimum at a value of 
$\rho$ slightly higher than the error in the ground state  $\delta E_0$ 
The consistency of the two gaps  $\Delta_1, \Delta_0$ is very high 
and goes beyond that the error in the gap itself 
$\Delta_1-\Delta_{0}<\delta \Delta_1$ warranting that the 
spin rotational invariance of the system is preserved by the algorithm. 
The difference $\Delta_1-\Delta_0 = 5.3 \times 10^{-5}$ at $\rho=0.0$
reduces to  $1 \times 10^{-6}$ at $\rho=0.6$  
in the 2x10 ladder and from $1.2 \times 10^{-7}$ at $\rho=0.0$  
to  $1 \times 10^{-8}$ at $\rho=0.6$  in the 3x6 ladder.  
\begin{table}
\caption{\label{tab:table6}  Energy and gap $\Delta_1$ as a function of the ratio 
$\rho=J_{\perp}/J_{\parallel}$ for the 3x6 ladder with 
$J_\perp=0.1$.}
\begin{ruledtabular}
\begin{tabular}{|c|c|l|l|l|}
$3\times6$&$\rho$ &$E_G$&$\Delta_{1}$&$E_B$       \\ \hline
ED    &0.1&-0.41688122& 0.432623 & -.00128503     \\  \hline
2S-DMRG&0.1&-0.41687959& 0.43300687 &  -.00128340  \\ \hline
ED    &0.2&-0.41649678& 0.442359   &   -.00090059 \\ \hline
2S-DMRG&0.2&-0.41649591& 0.44260306 &  -.00089972  \\ \hline
ED    &0.3&-0.4161880 & 0.43300687 &  -.00059181   \\ \hline
2S-DMRG&0.3&-0.41618749& 0.45266032 &  -.00059130   \\ \hline
ED    &0.4&-0.41595617& 0.463113   &   -.00035998  \\ \hline
2S-DMRG&0.4&-0.41595588& 0.46317943 &  -.00035969 \\ \hline
ED    &0.5&-0.41580272& 0.474130   &  -.00020653 \\ \hline
2S-DMRG&0.5&-0.41580249& 0.47415897 &  -.00020630 \\ \hline
ED    &0.6&-0.41572872& 0.485568   &  -.00013253 \\ \hline
2S-DMRG&0.6&-0.41572856& 0.48559530 &  -.00013237  \\ \hline
ED    &0.7&-0.41573539& 0.484843   &   -.00013920 \\ \hline
2S-DMRG&0.7&-0.41573518& 0.48490468 &  -.00013899  \\ \hline
ED    &0.8&-0.4158235 & 0.471902 &    -.00022731 \\ \hline
2S-DMRG&0.8&-0.41582325& 0.47203771 &  -.00022706   \\ \hline
ED    &0.9&-0.4159940 & 0.459162   &     -.00022731 \\ \hline
2S-DMRG&0.9&-0.41599350& 0.45940760 &  -.00039731   \\ \hline
ED    &1.0&-0.4162475 & 0.446652   &   -.00039781 \\ \hline
2S-DMRG&1.0&-0.41624647& 0.44704608 &  -.00065028\\ 
\end{tabular}
\end{ruledtabular}
\end{table}



\begin{table}
\caption{\label{tab:table7}  Error in the ground state energy 
and exact binding energy for the 2x10 ladder at $\rho=0.0$ and 
$\rho=0.6$ for increasing values of 
$J_{\perp}$.} 
\begin{ruledtabular}
\begin{tabular}{|c|c|c|c|c|}
$2\times10$&\multicolumn{2}{c|}{$\rho=0.0$}&\multicolumn{2}{c|}{$\rho=0.6$} \\ \hline 
$J_{\perp}$&$\delta E_0$&$E_B^{\text{ED}}$&$\delta E_0$&$E_B^{\text{ED}}$ \\ \hline 
0.05&.0000173&-.0001815&.00000093&-.00002374\\ \hline
0.10&.0000715&-.0007880&.00000380&-.00009639\\ \hline
0.15&.0001695&-.0019401&.00000849&-.00022034\\ \hline
0.20&.0003229&-.0037922&.00001551&-.00039864\\ \hline
0.25&.0005477&-.0065235&.00002530&-.00063489\\ \hline
0.30&.0008634&-.0103230&.00003881&-.00093364\\ 
\end{tabular}
\end{ruledtabular}
\end{table}

\begin{table}
\caption{\label{tab:table8} Error in the ground state energy 
and exact binding energy for the 3x6 ladder at $\rho=0.0$ and 
$\rho=0.6$ for increasing values of 
$J_{\perp}$} 
\begin{ruledtabular}
\begin{tabular}{|c|c|c|c|c|}
$3\times6$&\multicolumn{2}{c|}{$\rho=0.0$}&\multicolumn{2}{c|}{$\rho=0.6$} \\ \hline 
$J_{\perp}$&$\delta E_0$&$E_B^{\text{ED}}$&$\delta E_0$&$E_B^{\text{ED}}$ \\ \hline 
0.05&.00000023&-.00042351&.00000003&-.00003234\\ \hline
0.10&.00000282&-.00174351&.00000016&-.00013256\\ \hline
0.15&.00001358&-.00402195&.00000052&-.00030640\\ \hline
0.20&.00004299&-.00729456&.00000127&-.00056073\\ \hline
0.25&.00010536&-.01156190&.00000275&-.00090401\\ \hline
0.30&.00021701&-.01678901&.00000554&-.00134668\\ 
\end{tabular}
\end{ruledtabular}
\end{table}

The method, being a perturbative renormalization approach,
looses some precision as we increase the perturbation. 
In the square lattice, the strength of the perturbation 
can be encoded in a single parameter of the Hamiltonian
$J_{\perp}$. Therefore, we observe that 
the precision reduces as $J_{\perp}$ increases. However, 
when we switch on the diagonal interaction $J_{d}$ for a fixed
value of $J_{\perp}$ the precision {\em increases}. Furthermore,    
results in tables \ref{tab:table5} and \ref{tab:table6} suggest 
that it is not the nominal value of the transverse couplings 
what limits the precision of this algorithm but the value 
of the binding energy, which is more directly related to the 
expectation value of the interchain coupling Hamiltonian. 
To illustrate this point we have compared values of
$\delta E_0$   at  
$\rho=0$ and $\rho=0.6$ (which is nearly  the value of 
$\rho$ with minimal binding energy). The results are presented 
in table \ref{tab:table7}  for the 2x10 system and in table    
\ref{tab:table8} for the 3x6. The precision is at least one order 
of magnitude higher in the  $\rho=0.6$ for both systems.

\subsection{Chains coupled in triangular arrangements}
An alternative way to couple the chain in a frustrated 
geometry is the anisotropic triangular lattice 
(see Fig \ref{ED-latt}). An additional motivation 
to study this system is the neutron 
scattering experiment in Cs$_2$CuCl$_4$  which 
suggests that this geometry and values of  
$J'_{\perp}=J_{\perp}=0.3$ are the essential 
ingredients for the model Hamiltonian. 
Although the maximum for binding energy in this model is about 
$J_{d}=J_{\perp}$, the qualitative behavior 
of all the magnitudes studied is the same than in the lattice with 
diagonal couplings. With the proposed 
physical realization of this model in Cs$_2$CuCl$_4$ in mind,
instead of presenting another extensive comparison 
of numerical data, we ask ourselves whether we can achieve 
acceptable precision in a theoretical study of that material.       
Therefore we focus in a value of $J_{\perp}=0.3$ 
and increase $J_{d}$ in a 3x6 ladder. The results 
are presented in table \ref{tab:table9}. The point 
$J'_{\perp}=J_{\perp}=0.3$ turns out to be optimal
because the errors are minimum in all the quantities
computed.

\begin{table}
\caption{\label{tab:table9} Ground state energy $E_{G}$, 
gap $\Delta_{1}$ and binding energy of the chains $E_B$ at
different values of $J'_{\perp}$ and $J_{\perp}=0.3$. 
The number of states kept is $m_{s1}=8, m_{s2}=32$ }
\begin{ruledtabular}
\begin{tabular}{|c|c|l|l|l|}
$3\times6$&$J'_{\perp}$ &$E_G$&$\Delta_{1}$&$E_B$       \\ \hline
ED    &0.0&-0.43238517&0.354879  &-.01678898\\ \hline
2S-DMRG&0.0&-0.43216816&0.36016502&-.01657197\\ \hline
ED    &0.1&-0.42595045&0.376078  &-.01035425 \\ \hline
2S-DMRG&0.1&-0.42604272&0.37907273&-.010480822\\ \hline
ED    &0.2&-0.42136694&0.404534  &-.00577075\\ \hline
2S-DMRG&0.2&-0.42132679&0.40566085&-.00573060\\ \hline
ED    &0.3&-0.41878783&0.441008  &-.0031642 \\ \hline
2S-DMRG&0.3&-0.41875861&0.44085353&-.00316242\\ \hline
ED    &0.4&-0.41876039&0.422128  &-.0031642 \\ \hline
2S-DMRG&0.4&-0.41871723&0.42255551&-.00312104\\ \hline
ED    &0.5&-0.42172544&0.322892  &-.00612925\\ \hline
2S-DMRG&0.5&-0.42158277&0.32995853&-.00598658\\ \hline
ED    &0.6&-0.42805333&0.235580  &-.01245714\\ \hline
2S-DMRG&0.6&-0.42751008&0.25335704&-.01191389\\ 
\end{tabular}
\end{ruledtabular}
\end{table}
\section{ Conclusions} 
\label{conclusions}
We have studied frustrated spin ladders using a 
recently proposed numerical renormalization 
group method for quasi-one-dimensional systems.
We have presented exhaustive comparison 
with exact diagonalization data, a test that any other 
numerical method in strongly correlated systems 
have had to pass. The two-step DMRG method 
has shown good performance in a large class 
of frustrated systems reaching in some cases 
a precision comparable with the 1D case. We find a close correlation 
between the binding energy of the chains 
and the accuracy in the computation of 
ground state energy and the gap.
We have also shown that a  
model of weakly coupled chains in a triangular 
geometry for Cs$_2$CuCl$_4$ is within the range of 
applicability of the method.

\section{Acknowledgments}  We would like to thank  Porscha McRobbie
for a critical reading of the paper and several suggestions.



\begin{thebibliography}{99}

\bibitem{MAT-moukouri} S. Moukouri. cond-mat/0312011 
\bibitem{TS1-moukouri} S. Moukouri and L.G. Caron, Phys. Rev. {\bf B 67},
                 092405 (2003).
\bibitem{TS2-moukouri} S. Moukouri.  cond-mat/0305608
\bibitem{Coldea} R. Coldea, D. A. Tennant, A. M. Tsvelik, and Z. Tylczynski
            Phys. Rev. Lett. {\bf 86}, 1335 (2001).
\bibitem{EVERTZ} H.G. Evertz, G. Lana and M. Marcu,
                 Phys. Rev. Lett. {\bf 70}, 875 (1993).
\bibitem{SANDVIK} A.W. Sandvik and J. Kurkij\"arvi, Phys. Rev. {\bf B 43},
                   5950 (1991); A.W. Sandvik, Phys. Rev. {\bf B 59},
                   14157 (1999).
\bibitem{DAGOTTO} E. Dagotto and A. Moreo, Phys. Rev. Lett. {\bf 63},
             2148 (1989). 
\bibitem{JONGH} M.S.L. du Croo de Jongh, M.J. van Leeuwen and W. van Saarloos
               Phys. Rev. {\bf B 62} 14844 (2000)
\bibitem{CAPRIOTTI} L. Capriotti and S. Sorella, Phys. Rev. Lett. {\bf 84},
                3173 (2000). L. Capriotti, F. Becca, A. Parola, S. Sorella
                Phys. Rev. Lett. {\bf 87}  097201 (2001).
\bibitem{WHITE} S.R. White, Phys. Rev. Lett. {\bf 69}, 2863 (1992). Phys.
              Rev. {\bf B 48}, 10 345 (1993).
\bibitem{DMRG-book} 'Density-Matrix Renormalization', Ed. By I. Peschel,
        X. Wang, M. Kaulke and K. Hallberg, Springer (1998)
\bibitem{BARNES} T. Barnes, E. Dagotto, J. Riera,E.S. Swanson, 
                  Phys. Rev. {\bf B 47} 3196 (1993)

\bibitem{SCIENCE-LAD} E. Dagotto and T.M. Rice. Science {\bf 271} 618 (1996).
\bibitem{DAVIDSON} E.R. Davidson, J. Comp. Phys. {\bf 17} 87 (1975)

\bibitem{ALLEN} D. Allen, F.H.L. Essler, A.A. Nersesyan.  
            Phys. Rev. B {\bf 61} 8871 (2000).   
     
\bibitem{TSVELIK} A.A. Nersesyan and A.M. Tsvelik, Phys. Rev. B {\bf 67},
              024422 (2003).




\end{thebibliography}
\end{document}